# Superconducting Gap Engineering in Tantalum-Alloy-Based Resonators

Chen Yang,[1, *] Faranak Bahrami,[1, *] Guangming Cheng,[2] Mayer Feldman,[1] Nana Shumiya,[1]
Stephen A. Lyon,[1] Nan Yao,[2] Andrew A. Houck,[1] Nathalie P. de Leon,[1] and Robert J. Cava[3]

[1]*Department of Electrical and Computer Engineering, Princeton University, Princeton, New Jersey 08540, USA*
[2]*Princeton Institute for the Science and Technology of Materials, Princeton University, Princeton, New Jersey 08540, USA*
[3]*Department of Chemistry, Princeton University, Princeton, New Jersey 08540, USA*

Utilizing tantalum (Ta) in superconducting circuits has led to significant improvements, such as high qubit lifetime ($T_1$) and quality factors in both qubits and resonators, suggesting that material optimization plays an important role in the development of superconducting circuits. Thus we here explore superconducting gap engineering in Ta-based devices as a powerful strategy for expanding the range of suitable host materials. By alloying 20 atomic percent (at.%) Hf into Ta thin films, we achieve a superconducting transition temperature ($T_c$) of 6.09 K as observed in DC transport measurements, reflecting an increase in the superconducting gap. We systematically vary deposition conditions to control film orientation and transport properties of Ta-Hf alloy thin films. We then confirm the enhancement in $T_c$ via microwave measurements at millikelvin temperatures. We verify the ~40% increase in $T_c$ relative to bare Ta devices, while the loss contributions from two-level systems (TLSs) and quasi-particles (QPs) remain unchanged in the low temperature regime. These findings emphasize the promise of material engineering in superconducting circuits and point to many potential material candidates for further exploration.

## I. INTRODUCTION

Superconducting qubits are among the most advanced platforms for quantum information, offering potential scalability and compatibility with standard device fabrication techniques [1, 2]. These systems have enabled key advances across a broad range of applications, including quantum computation [3–6], quantum simulation [7–11], and quantum error correction [12–15]. Among the materials investigated for superconducting qubits, Ta has emerged as a compelling candidate due to its self-terminating stoichiometric surface oxide and chemical resistance to harsh post-fabrication processes [16–21]. The surface forms a chemically stable stoichiometric $Ta_2O_5$ layer and is amenable to controlled thinning through surface treatments [22, 23]. These advantages allow body-centered cubic (BCC) $\alpha$-phase Ta to exhibit improved microwave performance under material optimization, and further advances in Ta-based qubit performance have indeed been closely tied to ongoing progress in material engineering [17, 24–26]. Buffered oxide etch (BOE) removes surface oxides and increases the quality factor, $Q_{int}$ [23], and noble-metal encapsulation deposited immediately after sputtering suppresses the stoichiometric oxide in Ta thin films [27, 28]. Recently, efforts to mitigate surface loss by treating the surface oxides and utilizing high-resistivity silicon substrates have led to Ta-based qubits with lifetime and coherence times exceeding 1 ms [17, 22, 23]. Guided by these successes in reducing dielectric loss in Ta-based devices, we aim to explore material engineering approaches that enhance the superconducting gap.

Alloying Ta with another element to enhance the superconducting performance expands the range of viable material candidates available to tackle different loss channels in superconducting circuits. Prior efforts have explored alloying Ta with elements such as nitrogen (N) and hafnium (Hf). Tantalum nitride (TaN) has attracted interest since it potentially leads to an oxygen-free surface [27, 29–39]. More recently, bulk studies have shown that doping Ta with 20 at.% Hf can raise the $T_c$ to approximately 6.3 K. However, the growth and properties of Ta–Hf thin films in the context of superconducting qubits remain unexplored [40].

Motivated by material advances in Ta-based alloy superconductors, we report the synthesis and material characterization of Ta-Hf alloy thin films. Structural analysis using X-ray diffraction (XRD) and scanning transmission electron microscopy (STEM) (Fig. S4) verifies the BCC structure across all the films, and DC transport measurements reveal a 40% enhancement in $T_c$ relative to undoped $\alpha$-Ta ($T_c$ = 4.3 K) [41]. Surface characterization, including atomic force microscopy (AFM) and X-ray photoelectron spectroscopy (XPS), was employed to assess surface morphology and native oxide chemistry. AFM (Fig. S6) and STEM (Fig. S4) analyses further supported the dominant film orientation revealed by XRD. To evaluate microwave performance, coplanar waveguide (CPW) resonators were fabricated from a representative alloy film and measured at millikelvin temperatures (SI: Device fabrication). The resonator exhibited $Q_{int}$ comparable to that of undoped $\alpha$-Ta films, indicating that the microwave loss due to TLSs remains unchanged. Notably, a higher superconducting gap is observed during microwave measurements, as evidenced by the increased QP activation temperature, which is in agreement with DC transport. In terms of chemical resistance, the surface oxides protect the underneath alloy against aggressive acid treatment such as piranha (2:1 sulfuric acid to hydrogen peroxide) when removing hydrocarbon contamination on the surface [23?], while incorporation of Hf reduced the robustness of Ta against BOE as a trade-off.

Although Ta-Hf alloy thin films still require further optimization for surface oxide treatments, the untreated Ta–Hf films exhibit an enhanced superconducting gap and microwave performance comparable to native Ta , with no additional loss due to TLSs and QPs compared to $\alpha$-Ta. This study highlights the utility of alloying strategies for selectively tuning material properties relevant to superconducting circuits.

---

* These authors contributed equally to this work.

## II. SUPERCONDUCTING GAP ENGINEERING

Control over film growth parameters is essential to tailor structural and superconducting characteristics for thin films (SI: Film Deposition). In Ta-Hf alloy thin films grown by DC magnetron sputtering, both the crystallographic orientation and superconducting properties were tuned by adjusting the deposition duration and temperature with a similar deposition rate of ∼ 0.25 nm/s (Fig. 1). Overall, the Ta-Hf films exhibit considerably lower residual resistivity ratios (RRR) compared to undoped $\alpha$-Ta (a factor of ∼4 compared to dirty limit films [41]), which is expected as Hf incorporation introduces additional structural disorder in the form of smaller grain sizes (Fig. S6) and lattice mismatch. Our structural analysis shows that films with higher RRR values tend to adopt an out-of-plane (111) orientation, while films with smaller RRR values display mixed (110) and (111) orientations or are dominated by (110) (Fig. 1(b) and Table I). Furthermore, an increase in the deposition duration leads to a predominantly (110) orientation in the films (Table I). Additionally, we confirm that the Ta-Hf films possess an atomic composition of approximately 83%:17% Ta:Hf via elemental mapping analysis using STEM with energy-dispersive X-Ray spectroscopy (EDS).

Similar to the correlation between the thin film orientation and the RRR values from DC transport measurements, the structural evolution of Ta-Hf thin films also influences their $T_c$. Across all samples, $T_c$ values range between 6.09 ± 0.13 K (Table II). Films exhibit slightly lower $T_c$ as the deposition temperature is increasing. XRD analysis revealed that the full width at half maximum (FWHM) of the (111) peak decreased from 550 °C to 750 °C, indicating better crystallinity. However, at 850 °C, the structural peak broadened, suggesting a degradation in structural quality at higher temperatures. This trend suggests 750 °C as the optimum deposition temperature for achieving high crystallinity, supported by the DC transport results (Table II). These observations are mostly valid for films deposited over 1000 s (1 ks films), while the trend for films deposited over 2000 s (2 ks films) suggests a similar level of disorder for different deposition temperatures.

The interplay between structural variation and superconducting properties in Ta-Hf films is further influenced by lattice disorder and lattice mismatching. The lattice disorder within Ta-Hf alloy thin films is a result of the Hf doping. Hf has a slightly smaller atomic radius (∼ 208 pm) compared to Ta (∼ 200 pm), thus Hf doping expands the crystal lattice for all samples, which is confirmed by peaks shifting to lower $2\theta$ angles in XRD (Fig. 1(b) and Table I). This intrinsic mismatch brought the residual resistivity ($\rho$ ($\mu\Omega \cdot$ cm)) of the Ta-Hf alloy thin films to nearly five times that of the dirty-limit Ta films (Fig. 1(b) and Table II) [41]. Furthermore, deposition conditions lead to lattice mismatch between the film and the substrate. XRD and STEM analyses reveal that the lattice mismatch between the Ta-Hf alloy thin films and the sapphire substrate is approximately 1.6% for the 1 ks films, while the lattice mismatch increases to 20.5% for the 2 ks films (Fig. S4). Compared to the 1.4−1.9% mismatch observed in undoped Ta (111) orientation films [41], the 1 ks films show a lattice mismatch between the clean- and dirty-limit Ta films, while the 2 ks films show a significantly larger lattice mismatch. This feature appears as a blurry atomic structure in the STEM image of the 2 ks film (Fig. S4). Both lattice disorder and lattice mismatch contribute to increased residual resistivity and decreased RRR, emphasizing the critical role of lattice compatibility in optimizing transport properties [39].

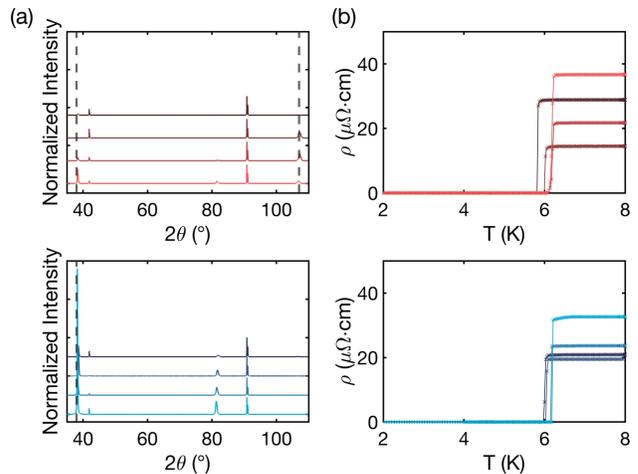

FIG. 1: **Structural and DC Transport Characterizations.** (a) XRD patterns of Ta-Hf alloy thin films sputtered at varying deposition temperatures and coating durations. The top panel (red tones) shows data from films coated for 1000 s (1 ks films), and the bottom panel (blue tones) corresponds to films coated for 2000 s (2 ks films). Within each panel, the darkest to lightest traces represent films deposited at 850 °C, 750 °C, 650 °C, and 550 °C, respectively. The dashed vertical lines indicate the expected peak positions for the (110) and (111) orientations, serving as a reference for identifying crystallographic orientation. A zoomed in plot of the (110) and (111) peaks of (a) are presented in Fig. 5. The XRD patterns for each deposition shifted along Y-axis for clear display. (b) Zero-field cooling DC resistivity of the films in (a) plotted as a function of temperature under zero applied magnetic field. Color mapping matches the corresponding deposition conditions and durations used in (a), illustrating how $T_c$ and residual resistivity vary with growth conditions.

## III. PARAMETERIZING MICROWAVE LOSSES

In superconducting microwave resonators, energy dissipation arises primarily from TLSs, QPs, trapped vortices, and power- and temperature-independent loss ($Q_{other}$) [23, 41]. TLSs are microscopic defects, located in amorphous dielectrics or at material interfaces in the resonator. These defects are particularly influential at low temperatures and single-photon microwave power, where they act as the dominant source of loss. The loss due to QPs becomes dominant at elevated temperatures and leads to an exponential decay in internal quality factor ($Q_{int}$) at higher temperatures (> 0.5 K). The slope of this decay is directly influenced by the superconducting gap.

We investigate the microwave loss behavior in the Ta-Hf thin films by measuring CPW resonators fabricated from this alloy, deposited at 750 °C for 2000 seconds. By conducting microwave measurements across varying temperatures and mi-crowave powers, we aim to evaluate different loss behaviors (e.g., TLSs, QPs, and vortices) compared to the bare Ta devices. Our findings reveal that at base temperature ($T \sim 20$ mK), $Q_{int}$ increases with rising microwave power, indicative of loss from saturable TLSs. At base temperature, $Q_{int}$ spans from approximately $1 \times 10^6$ at the lowest drive power to $4 \times 10^7$ at the highest; a similar range of $Q_{int}$ is observed in the Ta devices (Fig. 2(a)). Although the low and mid-temperature range shows similar behavior for both Ta-Hf and Ta devices [23], the activation temperature due to QPs is demonstrably increased in Ta-Hf resonators, suggesting that the superconducting gap is enhanced. Across seven measured resonators, the extracted $T_c$ for Ta-Hf resonators exhibits a mean-value of $5.92 \pm 0.24$ K, consistent with DC transport measurements (Fig. 1(b) and Table II). Additionally, we compare the extracted amplitude proportional to the kinetic inductance ratio ($A_{qp}$) for Ta-Hf and bare Ta devices, and we confirm that the loss due to the QPs remains comparable to the dirty-limit Ta devices while the $T_c$ is increased by 40% (Fig. S5) [23, 41]. To quantitatively assess TLS-induced losses in the Ta-Hf films, we extracted the inverse linear absorption from TLSs, ($Q_{TLS,0}$), and extracted the surface loss for measured Ta-Hf resonators. The corresponding surface loss tangent (tan $\delta$) for the Ta-Hf alloy is $(16.9 \pm 1.0) \times 10^{-4}$, which within the uncertainty overlaps the surface loss tangent of $(15.9 \pm 0.7) \times 10^{-4}$ measured for native Ta (Fig. 2(b)) [23]. The trend agrees well with reference data from native Ta resonators [23], indicating that Hf incorporation does not introduce additional TLS loss and suggesting that alloying with Hf preserves dielectric performance.

TABLE I: Fitted parameters of XRD peaks belonging to Ta-Hf alloy 1 ks thin films in Fig. 1(a). Values are relative to the substrate peak at $2\theta \approx 90°$.

| Sputtering Temp. (°C) | (110) Peak | | | (111) Peak | | | Lattice constant (Å) |
|---|---|---|---|---|---|---|---|
| | Location (°) | FWHM (°) | Normalized Intensity | Location (°) | FWHM (°) | Normalized Intensity | |
| 550 | 37.95 | 0.371 | $0.307 \pm 0.007$ | 106.91 | $0.851 \pm 0.015$ | 0.15 | 3.305 |
| 650 | 38.36 | 0.415 | $0.328 \pm 0.005$ | 107.00 | $0.731 \pm 0.016$ | 0.49 | 3.309 |
| 750 | N/A | N/A | N/A | 107.08 | $0.725 \pm 0.017$ | 0.50 | 3.312 |
| 850 | 38.40 | 0.814 | $0.589 \pm 0.007$ | N/A | N/A | N/A | 3.316 |

TABLE II: Fitted parameters for resistivity measurements of Ta-Hf alloy thin films with different coating times and sputtering temperatures. Residual resistivity ($\rho$) and RRR are estimated from full-range $\rho$ vs. $T$ measurements. The error in $T_c$ is defined as the data point collection increment measured by PPMS.

| Coating Time (s) | Sputtering Temperature (°C) | $T_c$ (K) | RRR | Residual Resistivity ($\mu\Omega \cdot$ cm) |
|---|---|---|---|---|
| 1000 | 550 | $6.17 \pm 0.01$ | $1.53 \pm 0.01$ | $36.61 \pm 0.01$ |
| | 650 | $6.20 \pm 0.01$ | $1.45 \pm 0.01$ | $21.72 \pm 0.01$ |
| | 750 | $6.02 \pm 0.01$ | $2.21 \pm 0.02$ | $14.48 \pm 0.01$ |
| | 850 | $5.84 \pm 0.01$ | $1.82 \pm 0.01$ | $28.85 \pm 0.02$ |
| 2000 | 550 | $6.21 \pm 0.01$ | $1.71 \pm 0.01$ | $32.55 \pm 0.02$ |
| | 650 | $6.18 \pm 0.01$ | $1.58 \pm 0.01$ | $23.64 \pm 0.02$ |
| | 750 | $6.06 \pm 0.01$ | $1.81 \pm 0.01$ | $20.88 \pm 0.02$ |
| | 850 | $6.00 \pm 0.01$ | $1.91 \pm 0.01$ | $20.89 \pm 0.02$ |

These results confirm that the dominant low-temperature dissipation in the Ta-Hf alloy resonators originates from saturable TLSs, and that Hf doping does not compromise the microwave performance at conditions relevant to qubit performance. However, its weak chemical resistance to BOE limits further improvements in the surface oxide treatments for these devices (Fig. S7). Furthermore, we do not observe loss due to vortex motion since the Ta-Hf alloy thin films intrinsically contain more defects which pin down the vortices and suppress the vortex-motion–induced loss [41, 42].

## IV. SURFACE OXIDE CHEMISTRY IN THE TA-HF ALLOY

To further investigate the dielectric loss limiting the performance of Ta-Hf alloys relative to native Ta, we employ XPS to characterize the surface oxide in our Ta-Hf thin films (SI: XPS Analysis). Core-level spectra were collected for the Ta and Hf 4$f$ electrons to identify the chemical states present on the film surface (Fig. 3). For Ta, the metallic doublets appear in the 20–24 eV range and the Ta$^{5+}$ doublets are observed between $25 - 29$ eV which align with those reported for native Ta (Fig. 3(a)) [22, 43, 44]. Suboxides of Ta, specifically Ta$^{1+}_{7/2}$ and Ta$^{3+}_{7/2}$, are fitted at 23 eV and 24 eV, respectively, which fall within the reported suboxides' peak range in native Ta. Hf oxides were analyzed similarly, although distinguishing individual suboxides is more challenging. The energy separation between the 4$f$ peaks for common Hf suboxides, such as Hf$^{2+}$ and Hf$^{3+}$, is less than 0.5 eV, making them difficult to resolve given the typical instrumental resolution of 0.1–0.2 eV. Therefore, we refined the Hf spectra using broader peaks attributed to fully oxidized HfO$_2$ and a less-defined HfO$_x$ state, following established Hf XPS analyses (Fig. 3(c)) [45, 46]. Reported literature finds the HfO$_2$ peak between 16–18 eV, while HfO$_x$ peaks appear around 14–16 eV, depending on the local chemical environment, with spin–orbit splitting values typically around $1.7 \pm 0.3$ eV. Our fitted Hf$^{4f}$ doublets between 16–18 eV align well with these reported values, as does the range for Hf$^{suboxides}_{7/2}$ between 14–16 eV [47] (Fig. 3(c)). Ad-ditional spectral shifts are observed due to different material interfaces. The Ta metal-oxide interface causes a shift in the Ta metallic doublets by approximately 0.4 eV, consistent with prior reports for native Ta films [22]. For Hf, similar metal-oxide interfacial component ($Hf^{int}$) shifted by 0.7 eV relative to the bulk metal peak [45, 47, 48]. We monitored oxide growth over an 18-day period following film deposition to better understand how Hf incorporation influences surface oxides. In both Ta and Hf spectra, the fraction of fully oxidized ($Ta_2O_5$ and $HfO_2$) states gradually increased during the 18-day exposure, indicating progressive surface oxidation (Fig. 3(b), (d)). To estimate the surface oxide thickness, we use cross-sectional EDS mapping (Fig. 4). The cross-sectional EDS mapping analysis suggests a 4 nm oxide thickness for the Ta-Hf alloy, thicker than the reported thickness for native Ta (3 nm) [22, 23]. In addition, we obtain the Ta-to-Hf atomic ratios for the bulk to be (83%:17%) via EDS analysis and the atomic ratio of Ta-to-Hf at the surface to be (78%:22%) via XPS analysis.

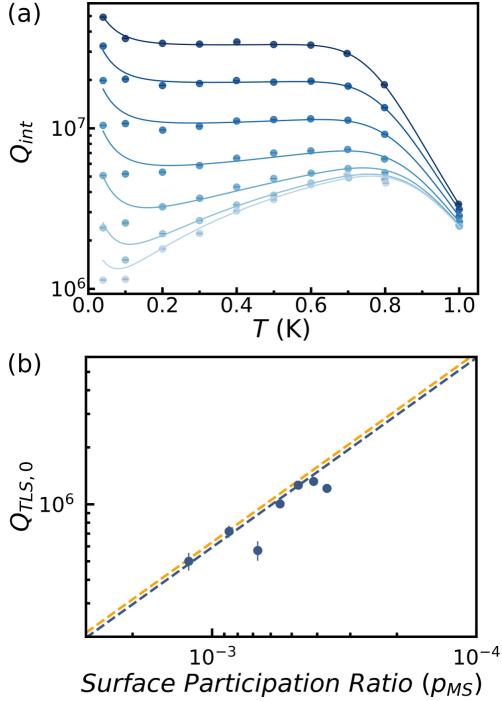

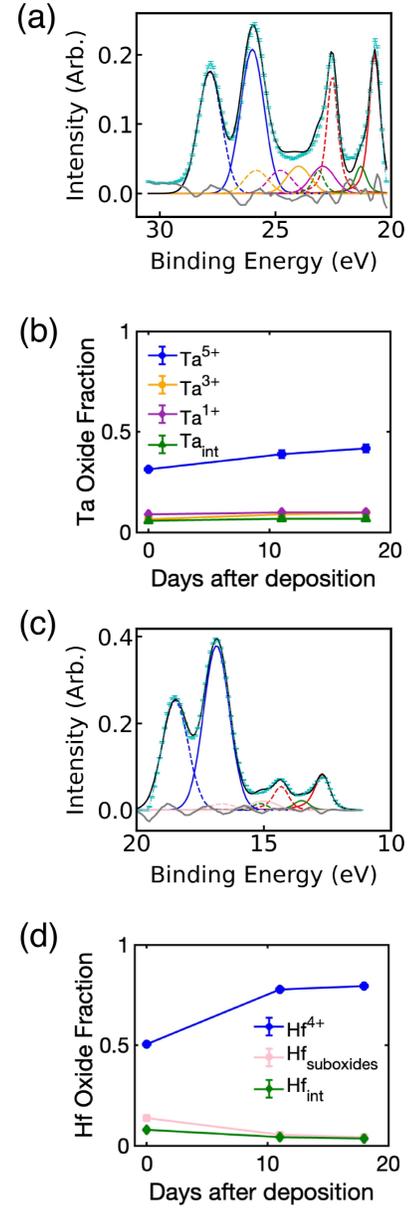

FIG. 2: **Microwave Losses in Ta-Hf Resonators**. (a) Temperature dependence of the internal quality factor ($Q_{int}$) measured at various microwave powers for a CPW resonator. At low temperatures and low powers, $Q_{int}$ is limited by saturable TLSs and rises with increasing power. At higher temperatures, an exponential decay in $Q_{int}$ indicates increasing losses from thermally activated QPs. The extracted $T_c$ for Ta-Hf resonators show a mean-value of 5.92 ± 0.24 K, confirming the enhancement in the superconducting gap. (b) Extracted $Q_{TLS,0}$ for seven Ta-Hf resonators plotted as a function of surface participation ratio ($p_{MS}$). The extracted surface loss tangent for Ta-Hf (blue dashed line) and Ta (orange dashed line, Ref. [23]) are $(16.9 \pm 1.0) \times 10^{-4}$ and $(15.9 \pm 0.7) \times 10^{-4}$, respectively. The shaded areas correspond to the uncertainties in the surface loss tangents.

FIG. 3: **Surface Oxide Evaluation over 18 Days of Exposure to Ambient.** The Ta (a) and Hf (c) binding energy spectra from XPS measurements. Both metallic and oxidation states exhibit a pair of peaks due to spin-orbit splitting. The metallic peaks show a sharp asymmetric shape, while the oxide peaks are modeled with Gaussian profile doublets. The metallic doublet peaks (red) are located at $11 - 14$ eV and $20 - 24$ eV for Hf and Ta, respectively. The atomic percentage for each oxide species and interface are presented over 18 days for Ta (b) and Hf (d). Both $Ta_2O_5$ and $HfO_2$ exhibit a gradual increase and saturation, while the $HfO_x$ shows an opposite trend.

## V. CONCLUSION

We demonstrate that the superconducting gap of Ta-based thin films can successfully be increased through alloying with Hf, achieving higher $T_c$ (6.09 K) without introducing additional microwave losses attributable to TLSs or QPs. Despite the thicker oxide and observable device surface roughness, the doping of Hf by Ta improved the chemical resilience of Hf against fabrication and post-fabrication processes, though surface oxide treatments such as BOE remain damaging to the material. Our work demonstrates how utilizing doping methods has the potential for tuning favorable material and fabrication features in the superconducting base layers of devices. This work also highlights the potential of material alloying as a versatile strategy for expanding the range of suitable host materials for superconducting circuits. Future efforts can focus on exploring alternative alloy systems, such as Ta-Zr [40] or other metal combinations, to further optimize material performance for superconducting circuits.

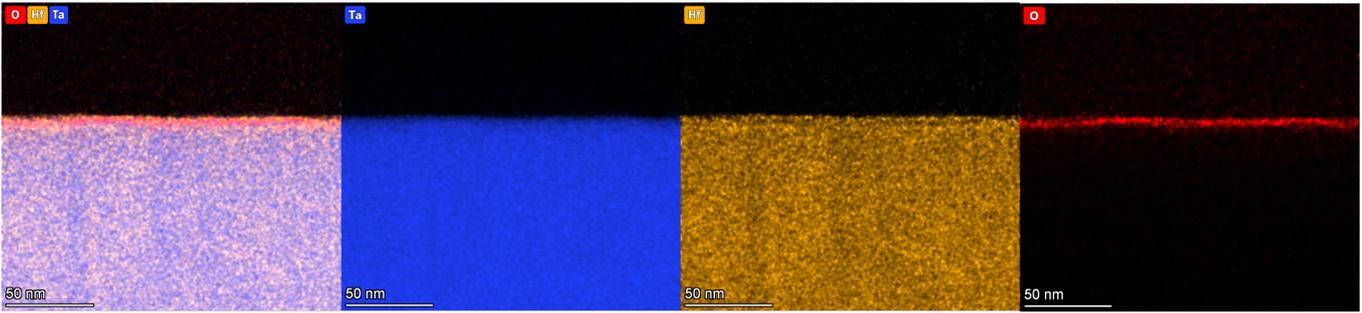

FIG. 4: **Cross-sectional EDS elemental mapping scan on Ta-Hf.** Cross-sectional EDS mapping Ta (blue) and Hf (orange) confirmed a uniform Ta-Hf alloy with the Ta to Hf atomic ratios for the bulk to be (83% : 17%). The oxygen (red) map reveals a surface oxide layer with the estimated surface oxide thickness of around 4 nm.

### SUPPORTING INFORMATION (SI)

The Supporting Information (SI) is available as a separate file. The SI includes detailed methods on film growth and device fabrication. In addition, discussions and results on Ta-Hf alloy as a Type-II superconductor are included. The observations in STEM cross-sectional images and surface morphology of Ta-Hf alloy thin films are available. We also presents the surface comparison before and after BOE (a mixture of ten parts 40% $NH_4F$ solution to one part 49% HF solution by volume) treatment. Lastly, a brief statement on XPS data analysis can be found.

### VI. ACKNOWLEDGEMENT

This material is based upon work supported by the U.S. Department of Energy, Office of Science, National Quantum Information Science Research Centers, Co-design Center for Quantum Advantage ($C^2QA$) under contract number DE-SC0012704. The authors acknowledge the use of Princeton's Imaging and Analysis Center, which is partially supported by the Princeton Center for Complex Materials, a National Science Foundation (NSF)-MRSEC program (DMR-2011750), as well as Princeton Mirco/Nano Fabrication Laboratory.

Princeton University Professor Andrew Houck is also a consultant for Quantum Circuits Incorporated (QCI). Due to his income from QCI, Princeton University has a management plan in place to mitigate a potential conflict of interest that could affect the design, conduct and reporting of this research.

## Appendix A: Film Growth

The Ta-Hf alloy thin films were epitaxially deposited onto c-axis HEMEX sapphire (Crystal Systems LLC, 99.99%). 3-inch wafers were diced into 1-inch squares to test different deposition conditions. Substrates were cleaned in freshly prepared piranha solution (2:1 sulfuric acid to hydrogen peroxide) for 20 minutes to remove organic contaminants, then sequentially rinsed in three deionized water baths to eliminate residual acids, dipped into IPA, and then dried under a pressurized nitrogen stream.

Deposition was performed using DC magnetron sputtering system (AJA International, Inc.) equipped with a single alloy target matching the desired atomic ratio of Ta and Hf. Prior to film growth, the sputtering chamber was heated to 50°C above the target deposition temperature for one hour to degas the chamber walls and sample holder as well as remove hydrocarbon contamination on the surface of substrate. The temperature was then reduced to the desired deposition value and held for an additional 20 minutes to achieve thermal equilibration and surface conditioning.

Film deposition commenced upon opening the shutter. Two deposition durations, 1000 seconds and 2000 seconds, were explored to investigate thickness-dependent effects on film orientation. Depositions were carried out at four temperatures: 550°C, 650°C, 750°C, and 850°C. The XRD and DC transport results (Fig. 1 and Fig. 5) both indicate that 750°C is the optimal deposition temperature. The measured growth rate was approximately 14 nm/min. Following deposition, the films were cooled naturally to room temperature.

## Appendix B: Device fabrication

To ensure a fair comparison of coherence between Ta and Ta-Hf alloy, the device fabrication process was intentionally kept the same as reported in Ref. [23]. Film structure was characterized using a Bruker D8 FOCUS diffractometer with Cu $K_\alpha$ radiation ($\lambda$ = 1.5406 Å). DC transport measurements were conducted using a Physical Property Measurement System (PPMS). Following characterization, the deposited film was coated with photoresist (PR) AZ1518 at 4000 rpm and soft baked at 95°C for 1 minute. The pattern was then written on the PR using a Heidelberg DL66+ with a 1.8 $\mu$m spot size. The PR was developed in AZ300MIF for 90 seconds and rinsed with deionized water for 30 seconds. During development and rinsing, the sample was carefully held and gently agitated to ensure consistent exposure to the developer. The pattern is transferred to the metal film by reactive-ion etching (RIE) with chlorine chemistry and the following parameters: chamber pressure of 5.4 mTorr, a radio frequency (RF) inductively coupled plasma (ICP) source of 500 W, a RF bias source of 50 W, with chlorine flow of 5 sccm and argon flow of 5 sccm.

The sidewalls of Ta-Hf alloy thin film devices were distinctly rougher compared to those of native Ta devices, showing porous features. Scanning electron microscopy (SEM) images captured after metal etching in chlorine gas for 100 seconds (Fig. 6). The sidewalls exhibit layered and porous morphologies, likely caused by the incorporation of Hf into the Ta lattice, which introduces strain and local variations in etching behavior. This roughness appears intrinsic to the alloy composition, since it persists for films with different deposition temperatures.

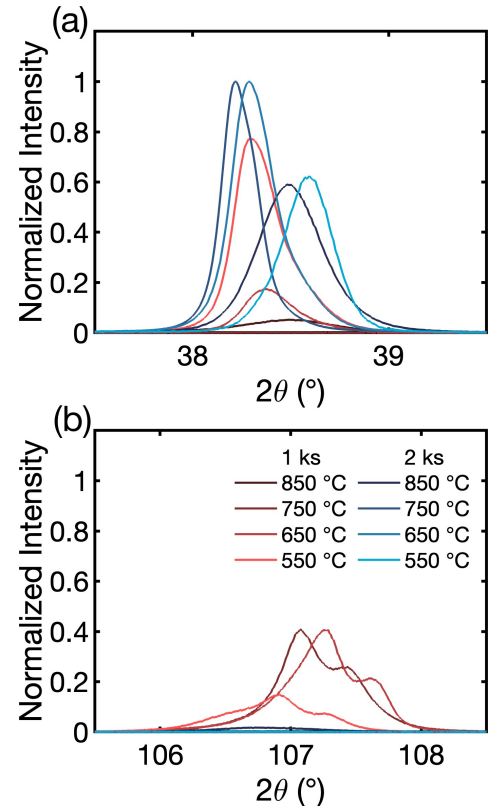

FIG. 5: Zoomed in XRD patterns of Ta-Hf thin films in Fig. 1. (a) Ta (110) peak ($2\theta \approx 38°$), showing peak shifts due to changes in lattice parameter with processing conditions. The peak shift among samples is approximately 0.8°. The 2000 s-coated samples overall have higher peak intensities as the (110) orientation is more predominant. (b) Ta (111) peak ($2\theta \approx 107°$) exclusively presents in 1 ks films. The curves show normalized intensities referenced to the substrate peak at $2\theta \approx 90°$.

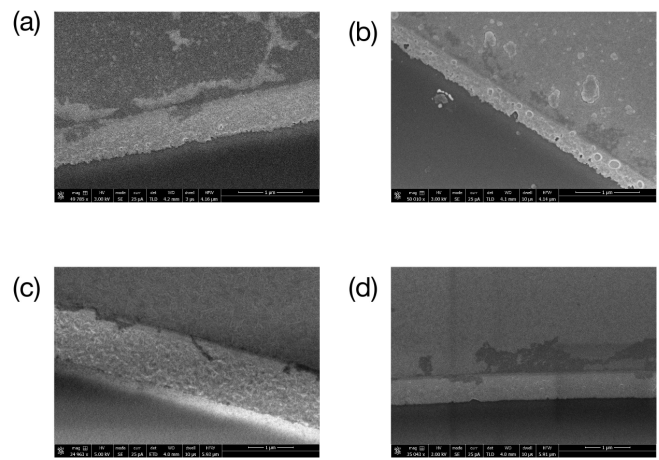

FIG. 6: SEM images of 1000s-coated Ta-Hf thin film sidewalls after etching in chlorine gas for 100 seconds. Panels (a)–(d) correspond to films sputtered at 550°C, 650°C, 750°C, and 850°C, respectively. Ta-Hf samples overall have porous and ridged sidewalls comparing to smooth sides walls of native Ta films [23].

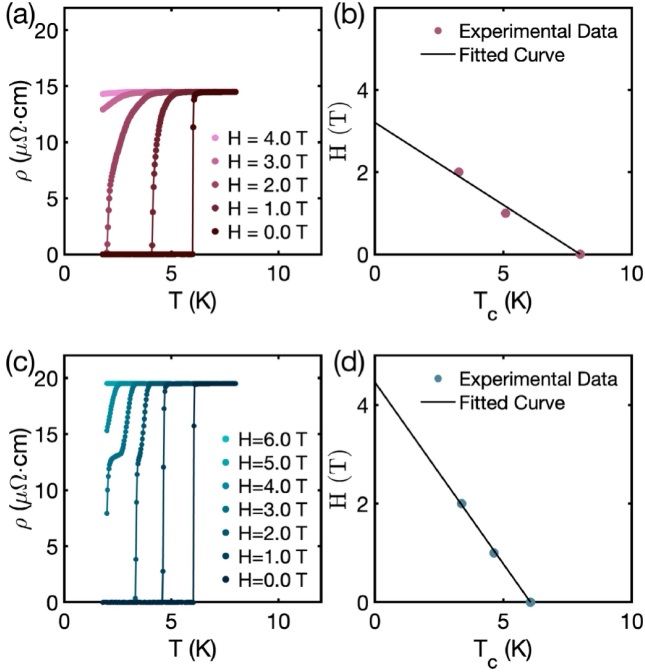

FIG. 7: Magnetoresistivity as function of temperature are shown for 1 ks (a) and 2 ks (c) films deposited at 750°C. The extracted $T_c$ at each applied magnetic field for 1 ks (b) and 2 ks (d) films are fitted using the WHH model, consistent with dirty-limit type-II superconductivity.

TABLE III: Superconducting parameters for Ta-Hf thin films deposited for 1000s and 2000s at 750°C. The upper limit critical fields ($H_{c2}(0)$) and extracted coherence lengths ($\xi(0)$) are presented.

| Deposition Temp (°C) | Deposition Time (s) | $H_{c2}(0)$ (T) | $\xi(0)$ (nm) |
|---|---|---|---|
| 750 | 1000 | 2.22 ± 0.20 | 10.86 ± 0.49 |
| 750 | 2000 | 3.54 ± 0.73 | 8.61 ± 0.89 |

To remove fabrication residue, the sample was cleaned with piranha solution followed three baths of DI water, Isopropyl Alcohol (IPA), and pressurized nitrogen stream. After cleaning, the chip was mounted on a commercial microwave package (QDevil QCage.24) with an associated Au-plated printed circuit board (PCB) similar to the packaging described in Ref. [23].

### Appendix C: Upper Limit Critical Field

The Ta-Hf thin films are dirty-limit type-II superconductors, characterized to have a upper critical field ($H_{c2}(0)$) and lower coherence length ($\xi(0)$) relative to native Ta. For this analysis, we focused on films sputtered at 750°C with coating times of 1000s (Fig.7(a),(b)) and 2000s (Fig.7(c),(d)).

To extract $H_{c2}(0)$, we traced the midpoint of the resistive transitions at each applied magnetic field and fitted the resulting data using the Werthamer–Helfand–Hohenberg (WHH) model, which is widely applied to dirty type-II superconductors for estimating the zero-temperature upper critical field and its temperature dependence [49]:

$$\mu_0 H_{c2}(0) = -0.693\, T_c \left.\frac{dH_{c2}}{dT}\right|_{T=T_c}$$

where $H_{c2}(0)$ denotes $H_{c2}$ at zero temperature. Ta-Hf alloy thin films exhibit an order-of-magnitude higher $H_{c2}(0)$ compared to dirty-limit native Ta films, resulting in significantly shorter $\xi(0)$ (Table III). The enhancement in $H_{c2}(0)$ arises from increased disorder introduced by Hf doping, which reduces the superconducting coherence length and strengthens vortex pinning [50, 51], supported by our fitted results (Table III). Additionally, the structural defects introduced by Hf doping is also likely facilitate vortex pinning, thus no vortex motion-induced losses were observed in the resonator measurements of Ta-Hf devices (Fig. 2(a)) [41].

### Appendix D: Transmission Electron Microscopy

We observed different coating durations under the same deposition temperature change film orientation preference, so we used STEM to examine two Ta-Hf thin films coated for 1000s (Fig. ??(a)) and 2000s (Fig. ??(b)) at 750°C. Two samples were cut along the c-axis of the sapphire substrates to show cross sections. Both films are continuous, uniform and polycrystalline. Additionally, no step change was observed in the 2000s-coating sample. Since (110) is more densely packed than (111) plane, a darkness transition would be observed if there is a film orientation switch. The bright-field image appeared to be consistent across the film, thus no orientation change in the 2000s-coating sample. The diffraction patterns (Fig. ??(c),(d)) confirmed BCC structure, supported by XRD results. EDS spectra (Fig. 4) were taken simultaneously with STEM images. The atomic ratio of Ta and Hf is closer to 83%: 17%.

### Appendix E: The correlation between $T_c$ and $A_{QP}$

$A_{QP}$ is an overall amplitude proportional to the kinetic inductance ratio and inversely proportional to loss due to QPs [23]. The $A_{QP}$ anti-correlates with $T_c$ (Fig. 8) and we observed a similar anti-correlation behavior as in Ref.[23]. Our extracted $A_{QP}$ values are comparable to the extracted $A_{QP}$ for dirty-limit native Ta.

### Appendix F: Surface morphology

Hf-doped Ta films exhibit overall rougher surfaces compared to undoped Ta films [41]. AFM characterization of the 1 ks films (Fig.9) reveals surface morphologies that are consistent with structural features observed in the corresponding XRD patterns (Fig.1 and Table I). The Ta-Hf thin films display higher surface roughness than undoped Ta films [22, 41]. Among the alloyed samples, the film annealed at 750 °C (Fig. 9(b)) exhibited the lowest root mean square roughness, with $R_q$ = 1.08 nm. The increased surface roughness in the Hf-doped films is attributed to the incorporation of Hf, whose slightly larger atomic radius compared to Ta, introduces additional lattice distortions and surface roughness.

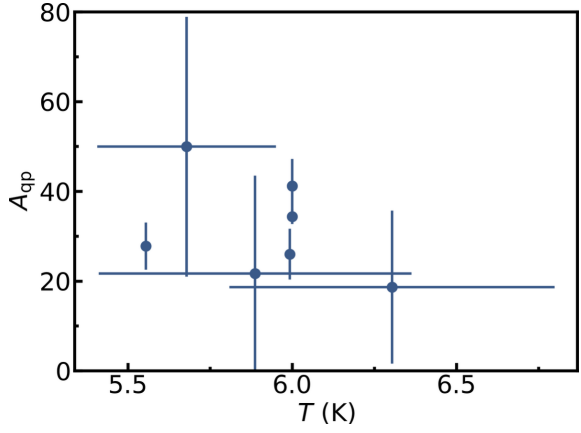

FIG. 8: The correlation between extracted $T_C$ and $A_{QP}$ for each measured Ta-Hf resonators. The extracted $A_{QP}$ are comparable to those obtained for dirty-limit Ta films, suggesting no additional loss due to QPs [23, 41].

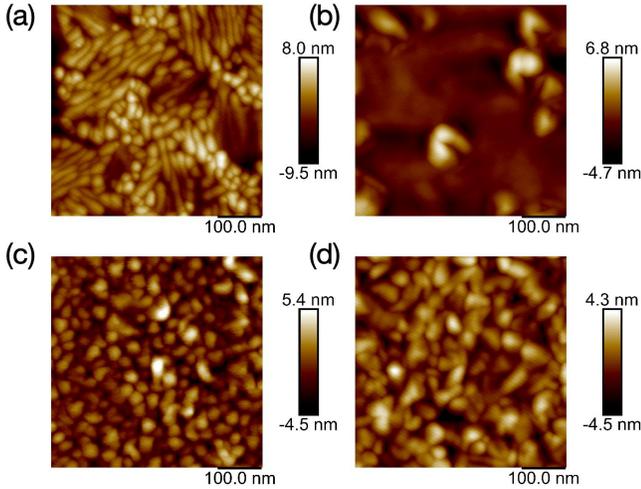

FIG. 9: AFM scans of the surface height of Ta-Hf thin films sputtered for 1000 seconds at 550 °C ($R_q$ = 2.86 nm) (a), 650 °C ($R_q$ = 1.36 nm) (b), 750 °C ($R_q$ = 1.08 nm) (c), and 850 °C ($R_q$ = 1.24 nm) (d). The surface roughness for Ta-Hf is comparable to the reported values for dirty-limit out-of-plane (110) Ta films (2.29 nm) and larger compared to dirty-limit out-of-plane (111) Ta films (0.543 nm) [41].

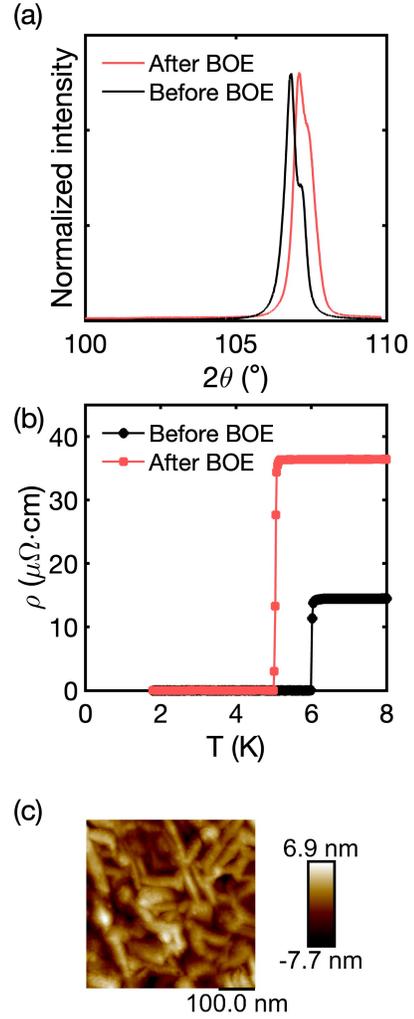

FIG. 10: Combined (a) structural, (b) DC transport and (c) surface morphological characterizations reveal low chemical rigidity of Ta-Hf alloy thin film under BOE treatment. The (222) Ta peak shows a shift to higher angle values, the $T_c$ decreased by 1 K from untreated to treated, and the $R_q$ worsened from 1.08 nm (untreated) to 2.0 nm (treated).

**Appendix G: Surface Chemical Resistance**

Although Hf doping to Ta improved the chemical resistance of Hf to aggressive acid cleaning treatment such as piranha, it reduced the chemical stability of Ta in BOE. The Ta-Hf thin films withstood a 20-minute aggressive acid cleaning exposure to piranha without showing any signs of bubbling or surface peeling. In contrast, the same Ta-Hf alloy thin film degraded rapidly in BOE. Visible peeling, wrinkling, and continuous bubbling occurred shortly after 1-minute-long immersion, indicating a strong chemical reaction between the film and the etchant.

XRD peak show broadening and shift to lower angle of the (222) Ta peak following BOE treatment (Fig.10(a)), pointing to structural changes. This was further supported by AFM images (Fig.10(c)), which revealed a rougher ($R_q$ = 2.0 nm) and more disordered surface. Electrical measurements showed that the $T_c$ dropped to 5 K (Fig. 10(b)), suggesting the possible loss of Hf atoms, which were responsible for the $T_c$ enhancement.

The observed degradation is likely due to the etching of the HfO$_2$ component in the surface oxide layer. While HfO$_2$ offers high chemical stability in oxidative environments, it is susceptible to be attacked by fluoride ions in BOE, which can break oxide bonds and expose the underlying Ta-Hf alloy layer to further damage. In contrast, the mixed oxide layer of Ta$_2$O$_5$ and HfO$_2$ appears to offer protection during aggressive acid cleaning treatment but not in fluoride-based enchants. Diluted HF is commonly used for wet etching HfO$_2$ thin film. Etched surface consistently show large surface roughness [52–58].

## Appendix H: XPS Analysis

The process of fitting XPS data to determine the chemical composition of surface oxide layers was adopted from Ref. [22]. The first step is to remove the background signal from inelastically scattered electrons using a Shirley background subtraction. Once the background is removed, the spectrum is refined into its constituent peaks, each representing a distinct chemical state. The peaks for metallic Ta or Hf (Ta$^0$ and Ta$_{int}^0$, Hf$^0$ and Hf$^{int}$) were fit using skewed Voigt profiles, while the oxide states (Ta$^{1+}$, Ta$^{3+}$, Ta$^{5+}$, Hf$^{suboxides}$, Hf$^{4+}$) peaks were fit with Gaussian profiles. The primary result from this fitting is the area under the curve for each peak, representing the photoelectron intensity ($A_n$) for that state. From these areas, the intensity fraction ($W_n$) for each oxidation state is calculated [22]:

$$W_n = \frac{A_n}{\sum_m A_m}$$